# ON TIME. VB: ELECTROMAGNETIC TIME*

C. K. Raju

*Indian Institute of Advanced Study, Rashtrapati Nivas, Shimla 171 005*

 ***Prefatory Note**: This paper first appeared in *Physics Education* (India) **9**, 1992, 251–265, as Part 5b, or the 7th in a series of 10 papers 'On Time'. These papers later also appeared as chapters in a book (*Time: Towards a Consistent Theory*, Kluwer Academic, Dordrecht, 1994). This is here being reproduced verbatim, except for this prefatory note. The various Parts cited in the text refer to the earlier papers/chapters. By way of background, an important aim of this series/book was to explain how both relativity and key aspects of quantum mechanics follow from a better understanding of the nature of time in classical physics. Since some time has elapsed since this series of papers and book was published, this prefatory note also summarizes some subsequent developments.

This paper explains the advantages of dispensing with the field picture, in classical electrodynamics, and working directly with the functional differential equations (FDEs) that arise in the particle picture. A key claim of the paper is that FDEs immediately lead to physical consequences which destroy the 'Newtonian paradigm' — a term explicitly defined and related to ordinary differential equations (ODEs) on p. 5. A later book (*The Eleven Pictures of Time*, Sage, 2003), and other publications by this author, used the term 'instantaneity' — that the state at one instant of time, determines the states past and future. This applies also to (hyperbolic) partial differential equations (PDEs) but fails with FDEs, which permit history-dependence.

This author's claim — that the use of FDEs hence leads to a paradigm shift in physics — was earlier severely criticized by H. D. Zeh at an international meet at Groningen in 1999. Zeh argued that existing physics was good enough, and that no paradigm shift was needed or involved. This author responded that Zeh had incorrectly assumed that a fresh mathematical understanding of existing physics could not, by itself, bring about a radical change, especially in classical physics.

This debate was summarized in a subsequent publication (*Found. Phys.* **34**, 2004, 937–62, arxiv.org:0511235) where this author reiterated that while FDEs arise naturally in the particle picture, *and involve no new physical hypothesis*, they can nevertheless involve qualitatively novel physics, radically different from expectations based on the Newtonian paradigm. The paper asked how one set of physical principles could admit two 'fundamentally incompatible interpretations of instantaneity and history dependence'.

Briefly, FDEs are equivalent to a ***coupled*** system of ordinary and partial differential equations. FDEs arise in the pure particle picture, while the coupled system of ODEs and PDEs arises in the field+particle picture: in many-particle electrodynamics, the equations of particle motion, using the Heaviside-Lorentz force, are ODEs, which are coupled to Maxwell's equations which are PDEs. However, the typical treatment of the subject simply neglects this coupling, giving the illusion of instantaneity. Taking this neglected coupling into account allows for the qualitative properties of FDEs that are impossible for solutions of ODEs or PDEs alone. The 2004 paper (cited above) emphasized that FDEs are merely more *convenient* than a coupled system of ODEs+PDEs (with 'convenience' understood in the sense of Poincaré).

The special importance for physics, of these new qualitative features of FDEs, was brought out in Chapter 6b of the 1994 book which connected FDEs to axiomatic quantum mechanics. This abstract connection was made more intuitive in the 2004 paper (cited above) which numerically solved the 2-body FDEs for the classical hydrogen atom.

More recently, the use of FDEs was advocated by M. Atiyah, in his Einstein lecture of 21 October 2005, at the University of Nebraska-Lincoln, and in a subsequent lecture at the Kavli Institute of Theoretical Physics (http://online.kitp.ucsb.edu/online/strings05/atiyah/). Like this author, Atiyah too claimed that the use of FDEs leads to a paradigm shift in physics. The similarity went a step further, for the relation between FDEs and quantum mechanics which was established as a theorem in this author's 1994 book, and by calculation in this author's 2004 paper (cited above), was proposed by Atiyah in 2005 as a conjecture.

However, in his Einstein lecture and talk at the Kavli Institute, Atiyah did not refer to this author's prior papers or two books, and instead explicitly said he was the first to suggest this line of thought. He was quoted to this effect in a subsequent article (http://www.ams.org/notices/200606/comm-walker.pdf,*Notices of the AMS*, **53**, 2006, 674–78), which named this as "Atiyah's hypothesis": that physics should use FDEs instead of ordinary differential equations (ODEs) or partial differential equations (PDEs). That article reiterated the potential of FDEs to bring about a paradigm shift in physics, but this author's work again went unacknowledged.

This raised some peculiar ethical issues, especially since Atiyah had already been explicitly informed earlier of this author's work — as early as 26 October 2005 — and had then responded to the author (Atiyah, personal communication). Further, an author of the 2006 article in the *Notices of the AMS* (M. Walker, personal communication) confirmed the natural assumption that a draft of this article, reporting Atiyah's Einstein lecture, was shown to Atiyah before submission. This author's prior work connecting FDEs to quantum mechanics was acknowledged only a year later (*Notices of the AMS*, **54**, 2007, p. 472, http://www.ams.org/notices/200704/commentary-web.pdf). Setting aside the ethical issues, which have been discussed elsewhere, and also partly investigated and reported on its website by the Society for Scientific Values (http://scientificvalues.org/cases.html), there is a key point about physics here.

The use of FDEs in physics requires no special hypothesis. As clarified above, and in great detail in this author's 2004 paper, FDEs arise naturally in physics if we take into account the neglected coupling of PDEs and ODEs as in Maxwell's equations coupled with the Heaviside-Lorentz force law, both of which are part of long-existing physics. Therefore, it is a conceptual mistake to advocate the use of FDEs as a new hypothesis, as was done by Atiyah during his lectures, and as is suggested by the unfortunate terminology of "Atiyah's hypothesis". Although FDEs arise naturally in existing physics, they have been a source of confusion for nearly a century due to the Newtonian paradigm — as the references cited in this paper show, a number of distinguished physicists took it for granted that history-dependent FDEs could be approximated by instantaneous ODEs, which is an incorrect procedure. Coming on top of this confusion, Atiyah's mistake in advocating the use of FDEs as a hypothesis is particularly unfortunate.

This mistake was indeed pointed out by this author to the *Notices of the AMS,* in a letter (http://ckraju.net/atiyah/Is_this_Ethical.pdf) which remains unpublished, for no clearly specified reason. Thus, Atiyah's mistake in promoting the use of FDEs as a new hypothesis still persists in a prominent and widely-read journal. This needs to be pointed out, else it may add to the already long-standing confusion about FDEs.

ABSTRACT. In the previous part we saw that the introduction of the field creates more problems than it solves. Here we begin by disregarding the problems due to the field, and formulate the two-body problem of retarded electrodynamics, without radiation reaction. The resulting equations of motion are time-asymmetric, and fail to satisfy the 'phase-flow' hypothesis underlying the recurrence and reversibility paradoxes. We present a counter-example to show that the Lorentz-Dirac equation, resulting in preacceleration, may be invalid since it is derived by replacing a retarded ordinary differential equation (o.d.e.) by a higher-order standard o.d.e. obtained by Taylor approximation. The solutions of advanced o.d.e.'s branch into the future, implying in-principle unpredictability from the past and resolving Popper's pond paradox. The branching and collapse of solutions of mixed o.d.e.'s suggests a resolution of the Wheeler-Feynman and grandfather paradoxes.

With a direct-action theory, or with Dirac's definition of radiation damping, the elimination of advanced interactions is a serious problem. We present an exposition of (i) the Sommerfeld condition, pointing out its arbitrariness; (ii) the Wheeler-Feynman absorber theory, pointing out its internal inconsistency; and (iii) the Hogarth-Hoyle-Narlikar theory, pointing out its external inconsistency. The remaining absorber theory predicts the existence of rare advanced interactions. We compare this with the empirical results of Partridge, and suggest that experiment proposed by Heron and Pegg may now be revived.

## 1 Introduction

**P**art IV ended with the hope that the introduction of the field may help to resolve the paradoxes of thermodynamics. But in Part VA we saw that the introduction of the field seems to create more problems than it solves.

Let us, for the moment, forget about the field, or else let us adopt the point of view that the field is a dispensable intermediary between particles. Let us look, instead, at the nature of the many-body problem (of electrodynamics), and the novel features arising from the *finite speed of interaction*: to measure time, or anything else, one must postulate that the speed of light is constant.

What are the consequences of the finite speed of interaction? We recall Poincaré's remark: '*The state of the world will depend not only on the moment just preceding, but on much older states.*' In both cases we obtain what Poincaré called '*equations of finite differences*'. As Poincaré further argued, the substance of physics lies in its mathematical formalism — the 'mechanical explanations' are redundant. So, between fields and particles, it matters little which mental picture we feel comfortable with.

How does the finiteness of the speed of interaction affect the underlying equations? To see this, consider a system of *n* particles. In the field picture, an accelerated charged particle $e_1$ gives out retarded radiation which is incident upon other charged particles $e_2, e_3, ..., e_n$. The outgoing retarded wave accelerates other charged particles at *later* or retarded times. In the particle picture, only these accelerations matter: the acceleration of $e_1$ at time *t* depends upon the acceleration of $e_2, e_3, ..., e_n$, at past or retarded times, say $t-\tau_2, ..., t-\tau_n$. Ignoring, for the moment, the question of self-action and radiation damping, the difference between the field-picture and the particle-picture does *not* show up mathematically.

## 2  The two-body problem of electrodynamics

*2. 1  Formulation*

For example, the equations of motion of two charged particles $i$ and $j$, in *one* dimension, interacting solely through retarded radiation (and without radiative damping) take the form:[1]

$$\ddot{z}_i(s_i) = \frac{k}{m} \frac{[1+\dot{z}_i^2(s_i)]^{1/2} \; sgn\,[z_i(s_i)-z_j(s_j)]}{\left\{[z_i(s_i)-z_j(s_j)]\dot{z}_j(s_j) - |z_i(s_i)-z_j(s_j)| c\dot{t}_j(s_j)\right\}^2}, \tag{1}$$

where $(t, z)$, with the appropriate subscript, denote the coordinates of the world-lines of the particles, dots denote differentiation with respect to the proper times $s_i$, $s_j$ of the two particles, and $(i,j) = (1,2)$ or $(2,1)$. Given $s_i$, the *retarded proper time* $s_j$ in the above equation, corresponds to the point at which the backward null cone from the point $(t_i(s_i), z_i(s_i))$ meets the world line of particle $j$. This is obtained from (see Fig. 5):

$$c\left[t_i(s_i) - t_j(s_j)\right] = |z_i(s_i) - z_j(s_j)|. \tag{2}$$

The difficulty of having two independent variables can be removed by rewriting (1) as

$$\frac{v'_i(t)}{[1-v_i^2(t)/c^2]^{3/2}} = \frac{(-1)^i k}{m_i \tau_{ji}^2(t)} \cdot \frac{c-(-1)^j v_j(t-\tau_{ji}(t))}{c+(-1)^j v_j(t-\tau_{ji}(t))}, \tag{1'}$$

where $k = -e_1 e_2/c^2$, the explicit retardations

$$\tau_{ji}(t) = t - t_j(s_j) \tag{2'}$$

are obtained from equation (2), primes denote differentiation with respect to $t$, and $v_i = z'_i$ is the velocity.

Reforming the notation (to use bars rather than subscripts), and using units with $c = 1$, these equations may be rewritten:

$$\frac{v'(t)}{[1-v^2(t)]^{3/2}} = \frac{b}{\tau^2} \cdot \frac{1+\bar{v}(t-\tau)}{1-\bar{v}(t-\tau)}, \tag{3a}$$

$$\frac{\bar{v}'(t)}{[1-\bar{v}^2(t)]^{3/2}} = -\frac{\bar{b}}{\bar{\tau}^2} \cdot \frac{1-v(t-\bar{\tau})}{1+v(t-\bar{\tau})}, \tag{3b}$$

where $b = -k/m$, $\bar{b} = -k/\bar{m}$, and

$$\tau(t) = |x(t) - \bar{x}(t-\tau)|, \tag{4a}$$

$$\bar{\tau}(t) = |\bar{x}(t) - x(t-\bar{\tau})|. \tag{4b}$$

We can now see the difference quite clearly: *the character of the differential equation has changed.* The system of equations (3) is no longer a system of simple differential equations, but is a system of 'difference-differential equations' or a system of ordinary differential equations (o.d.e.'s) with *retarded deviating arguments*. The 'accelerations' of the particles at time $t$ depend upon their velocities $v$, $\bar{v}$ at past or retarded times, $t-\tau$, $t-\bar{\tau}$.

What is the significance of this change? To study the two-body problem of electrodynamics, one must study o.d.e.'s with deviating arguments: even the simplest qualitative features of such o.d.e.'s completely destroy the Newtonian paradigm, and suggest a resolution of the paradoxes of thermodynamics, and the paradoxes of advanced action.

*2.2 Some definitions*

A first-order o.d.e. has the form

$$x'(t) = f(t, x(t)), \tag{5}$$

where $f$ is some function (generally non-linear). It is well known that the most general system of o.d.e.'s can be reduced to a system of such equations, i.e., to the form (5), regarding $x$ as a vector, if necessary. It is also well known that (under some mild requirement of continuity on $f$) prescription of the 'initial value' $x(0)$, at an instant $t = 0$, determines a unique solution $x(t)$ of (5) in a neighbourhood $[-\delta, \delta]$ of $t = 0$. This is the Newtonian paradigm.

In contrast, a differential equation with deviating arguments has the form

$$x' = f(t, x(t), x(t-\tau)). \tag{7}$$

The 'dependent variable', the function $x$, appears for more than one value of its argument, the 'independent variable', $t$.

An equation with deviating arguments is classified as *retarded*, or *history dependent*, if the highest order derivative of the unknown function appears for exactly one value of the argument, and this argument is not less than all the arguments of the unknown function and its derivatives appearing in the equation. For example,

$$x'(t) = f(t, x(t), x(t-\tau(t))) \tag{8}$$

is called retarded if $\tau(t) > 0$.

Similarly, the equations of motion of charged particles interacting solely through advanced radiation correspond to *anticipatory* behaviour, or an o.d.e. with *advanced deviating arguments*,

$$x'(t) = f(t, x(t), x(t+\tau(t))). \tag{8}$$

The definition requires the same proviso as above, except that now the argument of the highest order derivative must be less than or equal to all the other arguments of the unknown function, i.e., $\tau(t)>0$ in (8), or $\tau(t)<0$ in (7).

More generally, a system of charged particles interacting through both advanced and retarded radiation displays *partly anticipatory* behavior, corresponding to an o.d.e. with *mixed-type* deviating arguments:

$$x'(t) = f(t, x(t), x(t-\tau_1(t)), x(t+\tau_2(t))). \tag{9}$$

From now on, equations of the type (7), (8), and (9) will be referred to as retarded, advanced, and mixed-type o.d.e.'s. The mathematical theory of such differential equations with deviating arguments, also known as functional differential equations, differs from the mathematical theory of the usual differential equations.

*2. 3 The recurrence paradox and the past-value problem*

Van Dam and Wigner[2] considered equations involving both retarded and advanced fields. They asserted (without proof) that instantaneous positions and velocities were sufficient to determine unique trajectories.

Now, with electromagnetic interactions taken into account, the many-body equations of motion (3) are retarded o.d.e.'s. However, for the simplest model of even a retarded differential equation, modeling a history-dependent situation, it is inadequate, in general, to provide initial data at a point. Consider, for instance, the o.d.e with constant retardation $\pi/2$,

$$x'(t) = x(t-\frac{\pi}{2}), \quad t \geq 0. \tag{10}$$

To obtain a unique solution, it is insufficient to specify only the state at one point of time, say $x(0)$. Thus, $x = \cos t$ and $x = \sin t$ are obvious solutions, and since the equation is linear $x = a \cdot \cos t + b \cdot \sin t$ is a solution for arbitrary constants $a$ and $b$, and both $a$ and $b$ cannot be determined from a knowledge of $x(0)$.

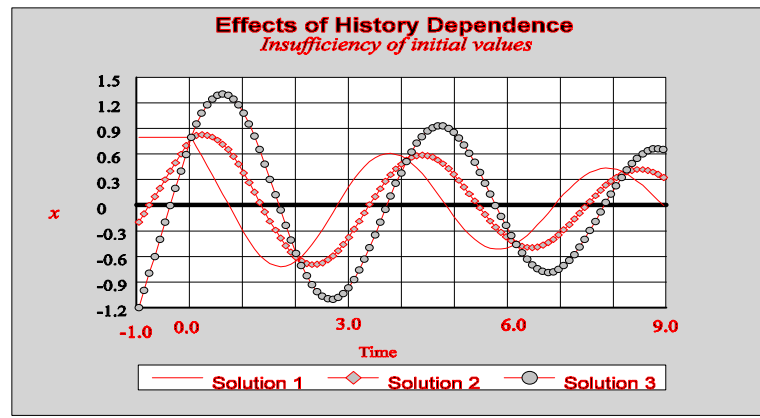

**Fig. 1: Effects of history-dependence**

*Three solutions of a model first-order delay equation, showing the effects of history-dependence. All three solutions have the same initial value x(0), though the past histories prescribed over the time-interval [-1, 0] are different. Note also the discontinuity in the derivative of solution 1, visible at t=0.*

Since the behavior is history-dependent, it is more reasonable to ask for a unique solution after prescribing the *past history*, i.e., an *initial function* $x = \varphi$, over the relevant part of the past: the interval of retardation, $[-\pi/2, 0]$.

In general, a unique solution of the past-value problem for the retarded system,

$$x'(t) = f(t, x(t-\tau_1(t), t-\tau_2(t), ..., t-\tau_n(t))), \qquad (11)$$

may be obtained under the following sufficient conditions.[3] (i) All delays, $\tau_i$, are bounded, and (ii) some technical conditions such as a local Lipshitz condition and a continuity condition are satisfied.

From the point of view of thermodynamics, the interesting conclusion is the following. The hypotheses underlying the recurrence paradox have been destroyed: *there is no longer a unique trajectory through each point of phase space*. More than one trajectory may pass through each point of phase space; trajectories may intersect (Fig. 1).

*2. 4 The reversibility paradox: time asymmetry of delay*

For retarded o.d.e.'s the intersection of trajectories takes place preferentially towards the future, in a way that destroys the hypothesis underlying the reversibility paradox. An ordinary differential equation is time symmetric: it may be solved either forward or backward in time. From a knowledge of the current values, Newton's laws may be used to predict the future or retrodict the past. However, a retarded o.d.e. relates past causes to current effects. Such an equation may be solved forward in time, but not, in general, backwards in time.

Consider the following ordinary, linear, retarded differential equation with constant coefficients, and constant retardation $r$:

$$x'(t) = a\,x(t) + b\,x(t-r), \qquad (12)$$

with $b$ different from zero and $r>0$. To solve the equation backwards, it is only necessary to solve an algebraic equation,

$$x(t-r) = \frac{x'(t) - a\,x(t)}{b}, \qquad (13)$$

to obtain the solution on $[t-2r, t-r]$, given $x = \varphi$ on $[-r, 0]$. For nonlinear equations, this already means that backwards solutions will not be unique. For the case under consideration, suppose $a \neq -b$ and we prescribe $\varphi(t) \equiv k$, a constant, on $[-r, 0]$, and ask for a backwards solution for $t \leq 0$. Then (13) implies that $x(t) = -ak/b$ so that the unique solution of the algebraic equation (13) fails to be continuous, and hence differentiable. Therefore, a (continuous) backwards solution of (12) does not exist in general.

Of course, one could think of choosing a final function in such a way that the solution exists. But then the solution would, in general, fail to be unique. Consider

$$x'(t) = b(t)\,x(t-1), \qquad (14)$$

where *b* is any sufficiently smooth (e.g. continuous) function which vanishes outside [0, 1], and with

$$\int b(t)\, dt = -1. \tag{15}$$

For example,

$$b(t) = \begin{cases} 0 & t \leq 0, \\ -1 + \cos 2\pi t & 0 \leq t \leq 1, \\ 0 & t \geq 1. \end{cases} \tag{16}$$

For $t \leq 0$, (14) reduces to $x'(t) = 0$ so that, for $t \leq 0$, $x(t) = k$ for some constant $k$. Now if $k$ is *any* constant then, for $t \in [0,1]$,

$$\begin{aligned} x(t) &= x(0) + \int_0^t x'(s)\, ds \\ &= x(0) + \int_0^t b(s)\, x(s-1)\, ds \\ &= x(0) + x(0) \int_0^t b(s)\, ds, \end{aligned} \tag{17}$$

since $x(s-1) \equiv k = x(0)$ on [-1, 0]. Hence, using (15), $x(1) = 0$ no matter what $k$ was. But $x(1) = 0$, and $b(t) = 0$ for $t \geq 1$, implies, by (14), that $x(t) = 0$ for *all* $t \geq 1$. Consequently, (14) does not admit a unique backwards solution even if we prescribe future data for all future times $t \geq 1$. Thus, if $\varphi$ differs from 0 on $[1,\infty)$ there are no backward solutions. But, if $\varphi \equiv 0$ on $[1, \infty)$, the solutions branch into the past (Fig. 2), and there is no way to pick a unique solution from the infinity of continuous solutions that are available.

### 2. 5  Preacceleration: the Taylor-series approximation

We saw in Part VA that the study of radiative damping and, in particular, the Schott term, leads to equations that are of the third order in time, resulting in the preacceleration of the electron. Dirac, in 1938, obtained these equations by means of a Taylor expansion which seems unavoidable.[4] Many other authors[5] have attempted similar approximation procedures, using a Taylor series expansion to get rid of retarded/advanced expressions, in dealing with the two-body problem in electrodynamics and gravitation. Physically, this procedure means that we model a history-dependent system by an instantaneous system with additional degrees of freedom.

   This procedure is known to be, in general, invalid. This may be seen from the following counter-example,

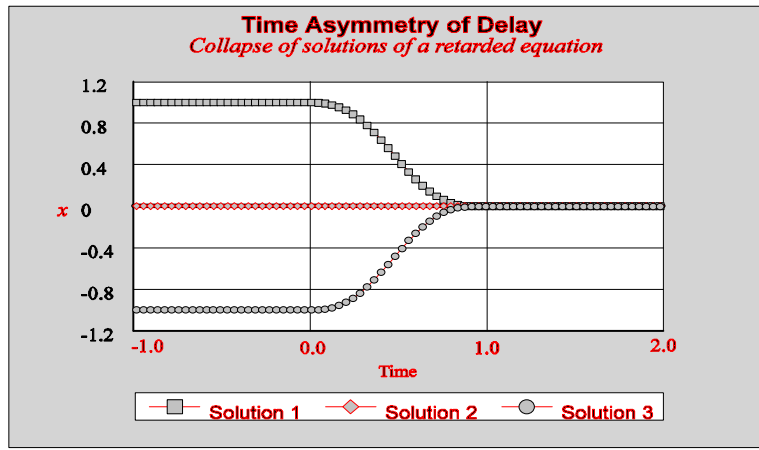

**Fig 2: Time asymmetry of delay**

*Three solutions of the retarded equation (14) which collapse towards the future. The different past histories, presecribed over the time-interval [1,0], all result in the same future solution for t≥1. Retrodiction is hence impossible from future data prescribed over t ≥ 1. Teleological 'explanations' are impossible with history-dependent evolution.*

$$x'(t) = -2\,x(t) + x(t-r), \quad (18)$$

where $r>0$ is a small constant. Every solution of this equation is bounded[6] and tends to zero as $t\to\infty$. But if we choose the Taylor series approximation to the right hand side and truncate after two terms, we obtain

$$x'(t) = -2\,x(t) + [x(t) - rx'(t) + \frac{1}{2} r^2 x''(t)], \quad (19)$$

which admits exponentially increasing solutions $x(t) = c\exp(\alpha t)$, with $\alpha > 0$. Thus, the Taylor approximation of (18) by (19) leads to qualitatively incorrect behaviour, no matter how small $r$ is, so long as $r>0$.

It may be shown that it is not the order of the approximation which is at fault: with instantaneous data, even an infinite number of degrees of freedom is inadequate. The order of the approximation does, however, make a difference from the numerical point of view, as pointed out by El'sgol'ts,[7] 'since the transition is equivalent to the rejection of the term with the highest order derivative in an *unstable-type differential equation* with a small coefficient before the highest derivative.'[Emphasis mine]

In the usual treatment of the numerical solution of retarded o.d.e.'s, attention is focused upon the discontinuities that might arise at the ends of delay intervals (e.g. Solution 1 of Fig. 1). However, one would expect the general electrodynamic many-body problem to be 'stiff': there could be oscillations at widely varying frequencies. In view of the Dahlquist barrier,[8] A-stability fails for any rule higher than the trapezoidal rule, so that the Taylor approximation could be numerically misleading for derivatives of order greater than two.[9] Thus, Dirac was perhaps right in a way when he rejected the higher order terms as too complex to apply to 'a simple thing like the electron'.

To summarize, the origin of the Schott term in the Lorentz-Dirac equation of motion is mathematically dubious, and can result in qualitatively incorrect behaviour, though it may yet provide a more robust numerical approximation than would be obtained by the inclusion of higher-order relativistically covariant terms. The alternatives that have been proposed,[10] to the Lorentz-Dirac equation, have not proved satisfactory.[11]

*2. 6 The pond paradox: advanced equations*

Popper's pond paradox, considered in Part VA, also admits a simple resolution in the context. The paradox seeks to exclude advanced interactions on the grounds that 'without the aid of organization from the centre' it is impossible to arrange the coherence of generators required to produce a convergent ripple (advanced wave). The resolution is that advanced waves, or anticipatory phenomena, cannot be predicted from past information, and hence cannot be arranged.

The reasons are simple. In the first place, anticipation is as time-asymmetric as delay. The past-value problem for advanced o.d.e.'s is, in general, insoluble for exactly the same reasons that retarded o.d.e.'s do not admit backwards solutions (i.e., the future-value problem is insoluble for retarded o.d.e.'s).

As an explicit example, analogous to (14), consider the equation

$$x'(t) = b(t)\, x(t+1), \qquad (20)$$

where $b$ has the same properties as before, except that we now require

$$\int b(t)\, dt = 1 \qquad (21)$$

in place of (15). For example,

$$b(t) = \begin{cases} 0 & t \leq 0, \\ 1 - \cos 2\pi t & 0 \leq t \leq 1, \\ 0 & t \geq 1. \end{cases} \qquad (22)$$

For $t \geq 1$, $x'(t) = 0$ implies $x(t) \equiv k$, for some constant $k$. But, if $x(t) = k$ for $t \geq 1$, where $k$ is any constant, then, arguing in the same way that leads up to (17), we find

$$\begin{aligned} x(t) &= x(1) - \int_t^1 x'(s)\, ds \\ &= x(1) - \int_t^1 b(s)\, x(s+1)\, ds \\ &= x(1) - x(1) \int_t^1 b(s)\, ds, \end{aligned} \qquad (23)$$

since $x(s+1) \equiv x(1) = k$ for $s \in [0, 1]$. Hence, $x(0) = 0$, which implies $x(t) = 0$ for all $t \leq 0$, since $b(t) = 0$ for $t \leq 0$, so that, by (20), $x'(t)$, vanishes for $t \leq 0$.

Thus, the solutions of the advanced equation (20) may collapse towards the *past* in exactly the same way as solutions of retarded equations collapse towards the future.

Another way of looking at this is that solutions of the initial function problem for (20), with coefficient as in (22), *branch* into the future as shown in Fig. 3. Given the entire past history, φ, it is impossible to predict the future for an anticipatory system which obeys (20). Thus, if φ(*t*) is different from 0 anywhere on (-∞, 0], then (20) does not admit any solution. But if φ(*t*) ≡ 0 on (-∞,0] then (20) admits infinitely many continuous solutions, and there is

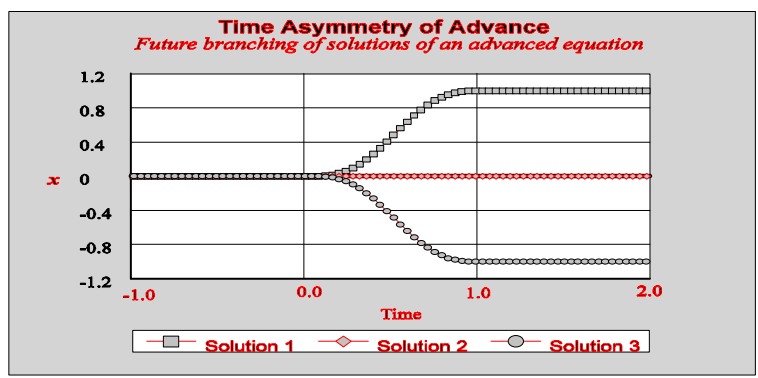

**Fig. 3: Time asymmetry of advance**

*Three solutions of the advanced equation (20) which branch towards the future. Three different future histories prescribed over the interval [1,2] result in the **same** solution for t ≤ 0. Prediction from past data, over t ≥ 0 is, thus, impossible. Causal 'explanations' are impossible for anticipatory phenomena.*

no way to pick a unique solution.

With retarded interactions, retrodiction is impossible; it would be absurd to rule out history-dependent phenomena on the grounds that all phenomena *ought* to be retrodictable. With advanced interactions, prediction is impossible; it is equally absurd to rule out anticipatory phenomena on the grounds that all phenomena *ought* to be predictable. It is the last requirement which leads to Popper's paradox.

*2. 7 Indeterminism: mixed deviating arguments*

Observational limits on advanced interactions, considered later on, show that the above kind of behaviour is not the norm. Nevertheless, it is possible to combine the two examples given above to get an idea of the type of situation that might prevail in a physically plausible universe in which both types of electromagnetic interactions are present, but advanced interactions form only a very small component, say one part in a trillion. In this case, the evolutionary equations would involve mixed deviating arguments.

No general methods are known for obtaining or even proving the existence of a solution of mixed-type o.d.e. But, the examples (14) and (20) may be combined to show that mixed-type o.d.e. will exhibit both the 'branching' and 'collapse' described earlier in Figs 2 and 3.

For example, consider

$$x'(t) = a(t) x(t-1) + b(t) x(t+1), \qquad (24)$$

where $b(t)$ has the properties as in (21) and (22). If we prescribe $x(t) = \varphi(t) \equiv 0$ for $t \leq 0$, then, on [0, 1], any solution of (20) is a solution of (24) since we have $x(t-1) = 0$ on [0, 1]. For $t>1$, $b(t) = 0$, so the equation to be solved is of type (14), $x'(t) = a(t)x(t-1)$, which shows that each (non-unique) solution of (24), on [0, 1], has a unique continuation for $t>1$.

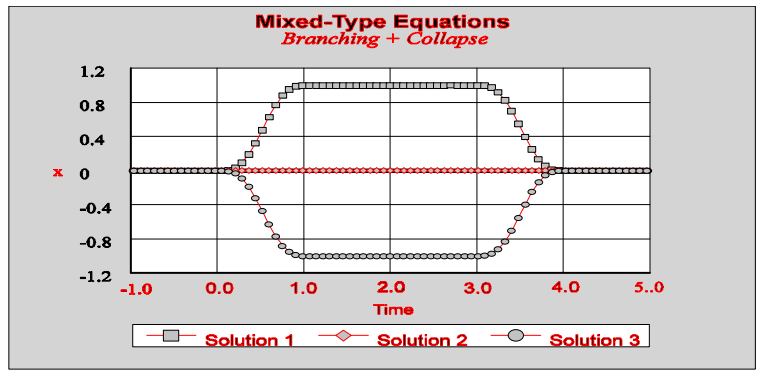

**Fig. 4: The mixed case**

*Three solutions of the mixed-type equation (24), which first branch and then collapse, showing that both prediction and retrodiction are, in general, impossible for such equations. Solutions of such equations may be regarded as intrinsically non-unique. The many-body equations of motion are of mixed-type with a small tilt in the arrow*

In particular, (24) does not provide a unique future solution even if the entire past history is prescribed. The opposite sort of behaviour, with branching into the past, and 'collapse' towards the future, is also possible. For instance, for (24), define $a(t)$ arbitrarily for $t \in (0, 1)$, $a(t) = 0$ for $t \in [1, 2]$, $a(t)$ continuous for $t \geq 2$ with

$$\int_2^3 a(t)\ dt = -1, \qquad (25)$$

and $a(t) \equiv 0$ for $t \geq 3$. Then the solutions of (24), for the same initial function $\varphi(t) \equiv 0$ for $t \leq 0$, are of the form shown in Fig. 4. This follows from the arguments given for equation (14) and in the above paragraph.

A couple of observations about mixed o.d.e.'s will illustrate the current state of knowledge. In the physics literature, precisely one exact solution is known, for the case of circular orbits,[12] and some numerical computations have been done for nearly-circular orbits.[13] From a general point of view, I am aware of only the works of Driver[14] and Schulman,[15] who recommends the use of boundary conditions in such a case and proposes a formal method of solution to the boundary value problem. The *kind* of result desired by Schulman was anticipated by Cooke and Krumme,[16] who relate the solution of an initial-boundary value problem for hyperbolic equations to the solution of a boundary value problem for a differential equation of mixed (neutral) type, describing the characteristics of the original hyperbolic system.

## 2. 8 *The Wheeler-Feynman paradox*

Wheeler and Feynman themselves proposed a resolution of their paradox by asserting that discontinuous forces do not exist in nature. I do not find this resolution satisfactory[17] for a paradox which is so well formulated.

One way out of this paradox is the well-known way of (hard) determinism. For example, the grandfather paradox may be deemed to have been resolved by the proof[18] that the Cauchy problem (initial value problem) is well posed in a spacetime with closed timelike curves. One might chance upon one's grandfather when he was a mere boy, but it would be mathematically impossible to kill him! We pointed out in Part I, the difficulty with this way: the empirical refutation of scientific theories tends to presuppose some indeterminism, or human freedom.

The other way out is to observe that the paradox actually demonstrates the *logical* incompatibility of the following three assumptions (stated informally):
(i)   one is free to perform any experiment,[19]
(ii)  interactions with the past are permitted,
(iii) the past is fixed and unchanging.
An additional assumption, which usually remains implicit (but was used in the preceding paragraph) is that
(iv) there is a univocal relationship between past and future (unlike the situation in Figs 3 and 4), even if interactions with the past are permitted.
The usual thing is to reject (ii) while ignoring the silent contradiction between (i) and (iv), and the less obvious contradiction between (ii) and (iv). We have shown that there is some coherence between (i) and (ii), though, if (ii) is assumed, one must give up (iv) and possibly (iii). A changing or 'dynamic' past would be very surprising, but it is not *logically* impossible; its physical validity must be decided by experiment! Current observational limits already show that advanced interactions, if they exist, must be very rare. In that case, it would be only very rarely, or at the microphysical level, that the past could be affected. We will examine this line of thought in greater detail in the next part, in the context of quantum mechanics.

## 3  The absorber theory of radiation

*3. 1  Action at a distance*

From the counter-examples in § 2.3, 2.4, and 2.5 above, it is clear that there are advantages to dispensing with the field picture, and adopting direct action at a distance. A theory based on retarded direct action would resolve the divergence difficulties of field theory as well as the paradoxes of thermodynamics. A theory based on mixed interactions may resolve even deeper problems.

There would be no great loss to intuition, because, as we saw in Part VA, the intuitive notions of 'contact' or 'locality' or 'chain of causes' are incorporated in the field concept in a way that is unintelligible or illusory. But the problem with the direct-action approach of Schwarzschild, Tetrode and Fokker was also pointed out: in order to balance action and reaction, one must have a time-symmetric interaction.

As we saw in § 2.6, 2.7, and 2.8 above, while advanced interactions do not lead to *logically* impossible situations, they imply situations that are physically fantastic, unless advanced interactions are very rare. The real world may not be perfectly time-asymmetric, but, locally at least, it is not time-symmetric. How should one reconcile the time-symmetric direct-action theory with observation?

*3. 2  The Sommerfeld radiation condition*

How is the reconciliation achieved in field theory? We saw, in Chapter VA, that the wave equation admits both retarded and advanced solutions; and, moreover, *any* field may be represented, within a certain volume, using both retarded and advanced fields:

$$F = F_{ret} + F_{out} = F_{adv} + F_{in}. \tag{26}$$

Both expressions describe the *same* physical situation.

In a famous paper[20] recording their disagreement, Einstein and Ritz observe that retarded and advanced fields are equivalent 'in some situations'. Ritz proposed that there should be a law against advanced fields! Einstein, on the other hand, thought that it was enough to restrict the solutions to be retarded: the exclusion of advanced fields was not law-like, but a matter of physical acceptability. Ritz regarded the choice of retarded fields as 'one of the roots of the Second Law [of thermodynamics],' while Einstein believed that 'irreversibility is exclusively based on reasons of probability.'

The analysis in Part IV and § 2 above shows that there are certainly merits to Ritz's point of view: it is easier to base irreversibility on retarded interactions than on reasons of probability. But the blanket exclusion of advanced fields would go against Dirac's covariant definition of the radiation field, which necessarily involves both advanced and retarded fields. Dirac adopted the pragmatic point of view that time-symmetric fields be accepted for all theoretical purposes, while, in practice, one could restore time asymmetry, by imposing the condition

$$F_{in} = 0. \tag{27}$$

This is known as the *Sommerfeld radiation condition*, in analogy with scattering theory, since it fixes the incoming wave. If (27) is imposed, the second equation in (26) fails.

The whole procedure seems a bit arbitrary: one uses advanced fields when convenient or necessary, and rejects them otherwise.[21] But even this arbitrary procedure is not available to the direct-action theory, for the field does not exist as an independent entity.

*3. 3  The Wheeler-Feynman theory*

Wheeler and Feynman[22] proposed the following procedure. The direct action theory of Schwarzschild, Tetrode, and Fokker is equivalent to field theory with time-symmetric fields. Given a collection of charged particles, the total field acting on a charged particle $k$ amounts to

$$\sum_{j \neq k} \frac{1}{2} \left( F_{(j)\ ret} + F_{(j)\ adv} \right). \tag{28}$$

The field that *ought* to act on particle $k$ according to retarded electrodynamics is

$$\sum_{j \neq k} F_{(j)\ ret} + \frac{1}{2} \left( F_{(k)\ ret} - F_{(k)\ adv} \right). \tag{29}$$

That is, the field acting on particle $k$ should consist of the retarded fields of all other particles together with only the radiation field of the particle $k$.

The necessary and sufficient condition for (28) and (29) to be equal is

$$\sum_{\text{all particles}} \frac{1}{2} (F_{ret} - F_{adv}) = 0. \tag{30}$$

Is there any physical circumstance in which (30) can be realized? If all the particles are shut inside an opaque enclosure, the absorber, we would certainly expect all radiation fields to vanish outside

$$\sum \frac{1}{2} (F_{ret} - F_{adv}) = 0 \quad \text{(outside)}. \tag{31}$$

But the radiation field has no singularities inside — it is a solution of the homogeneous wave equation — it must, therefore, actually vanish everywhere.

But complete destructive interference is impossible between converging and diverging fields. Therefore, we must also have

$$\sum F_{ret} = 0,$$

$$\sum F_{adv} = 0, \tag{32}$$

which is equivalent to the desired condition (30).

Thus, in a totally absorbing cosmos, time-symmetric interaction is equivalent to retarded interaction. A picturesque summary of the procedure is this: one sees the sun, or a distant star, just because its light is absorbed by the eye. The sun would not shine unless it knew that its light would eventually be absorbed. In this picture, which has been compared with Faraday's lines of force,[23] the source sends an 'offer' wave into the future, the absorber returns an advanced wave into the past, resulting in a 'transaction', or a giant handshake across spacetime.

Despite the appeal of the picture, the Wheeler-Feynman result is puzzling. How did time asymmetry emerge from time symmetry? Thus, inside a totally absorbing cosmos satisfying (32), we also have the other equality,

$$\sum_{j \neq k} \frac{1}{2} (F_{(j)\,ret} + F_{(j)\,adv}) = \sum_{j \neq k} F_{(j)\,adv} - \frac{1}{2} (F_{(k)\,ret} - F_{(k)\,adv}) = 0. \tag{33}$$

That is, fully advanced fields with radiative anti-damping are also consistent. Is there any reason to regard (29) as *more* valid than (33)? Wheeler and Feynman argued as follows. When particle $k$ is disturbed, the retarded radiation from the particle disturbs all other particles, at future times. The advanced (response) fields acting on the particle are therefore highly correlated. The retarded fields, due to past motions of absorber particles, may, however, be assumed to be uncorrelated. Therefore, on grounds of statistical mechanics, one must regard (29) as more valid than (33).

This argument rather begs the question. With time-symmetric interactions, one may no longer assume that particle motions are uncorrelated *before* interaction. Wheeler and Feynman gave other explicit calculations, which are also circular,[24] as are other similar arguments that have been advanced.[25]

*3. 4  Hogarth's theory*

Hogarth[26] pointed out that the Wheeler-Feynman calculation did not apply to an evolving universe. In the first place, the mean free path of a photon is of the order of a Hubble radius. That is, the radiation emitted by an antenna, in outer space say, would travel a long distance — to the very edge of the universe — before being absorbed. This would take a very long time, some 10 billion years. So, the advanced radiation would be absorbed some 10 billion years in the past, whereas retarded radiation would be absorbed some 10 billion years in the future. In an evolving universe, conditions are not likely to remain the same for such a long period of time, so the past absorber would be different from the future absorber.

Specifically, the absorption depends upon the (imaginary part of the) refractive index, which, in turn, depends upon the frequency of the radiation, and the density of charged particles. In an expanding cosmos, for instance, the density of charged particles in the past absorber would be much greater than the density of particles in the future absorber. Moreover, retarded radiation, traveling into the future would be red-shifted, while advanced radiation, traveling into the past would be blue-shifted. Finally, one can hope to compute the 'effective radius' of either absorber, without assuming it to be infinite, as was done by Wheeler and Feynman.

In view of these factors, the qualitative difference between the past and future absorbers may itself be the source of the asymmetry of radiation. Hogarth proposed a linear theory: the response $R$ of the past and future absorbers is a linear multiple of the incident stimulus fields $S$, so that

$$R_p = p\, S_p, \qquad R_f = f\, S_f, \tag{34}$$

where the *subscripts p* and *f* refer to the past and future, while the *coefficients p* and *f* are constants of proportionality. The stimulus field he took to consist of the radiation field from the particle, and the response from the other absorber.

The effective field of the particle is, therefore,

$$F_{eff} = \frac{1}{2} F_{ret} + \frac{1}{2} R_p + R_f. \tag{35}$$

From this follow the conditions for the consistency of retarded radiation: $f = 1$, $p \neq 1$. On the other hand, if $p=1$, $f \neq 1$, advanced radiation is consistent. If $p=f$, the radiation is time symmetric, no matter how close $p$ and $f$ are to 1. However, if $p=f=1$ the above equation becomes indeterminate, and leads to no conclusion.

The case $p=1$ corresponds to perfect absorption, or opacity, and holds for big bang cosmological models.[27] The case $f=1$ holds for models with a final singularity, or those which expand slower than the Dirac model. It has been argued that this case also holds for steady-state cosmologies.

Hoyle and Narlikar[28] propose a theory similar to Hogarth's, in that it demands an ideal future absorber, suitable for the steady state cosmology. Their theory differs in some details, however. In the first place they use an action principle in place of Hogarth's assumption of elementary time symmetric fields. Secondly, they prefer to base their theory on an argument very similar to the Wheeler-Feynman argument given above. Thirdly, they make a number of *ad hoc* hypotheses in the further development of the theory including its quantization; for example, they suppose, somewhat conveniently for the steady state theory, that the refractive index for advanced radiation behaves differently: advanced radiation 'builds up' the *further* it is from the source.

The Hogarth-Hoyle-Narlikar theory is, in any case, externally inconsistent: according to this theory, retarded radiation should be consistent in a steady state cosmos, while advanced radiation should be consistent in a big-bang cosmological model like the Einstein-de Sitter, and time-symmetric radiation in the closed Friedmann model. The microwave background radiation, discovered a quarter century ago, and interpreted as relic radiation from the big bang, seems to have sounded the death knell of the steady state theory.

*3. 5  Other theories*

Since the Wheeler-Feynman theory is internally inconsistent, this is not a very happy situation. I have proposed an alternative theory, which goes back to the notion of the electron as a finite-size object[29] to justify the hypothesis of a lower signal velocity for advanced radiation.

To calculate the absorber response, one now has to consider a countable infinity of stimulus and response fields, which can be summed to arrive at the following conclusions in terms of the response factors $p$ and $f$ for the past and future absorbers. Retarded radiation is consistent in those models which satisfy $p=1$ or $f=-1$ provided $|pf| \neq 1$, while advanced radiation is consistent in those models which satisfy $f=1$ or $p=-1$, $|pf| \neq 1$. If both $p=1$, $f=1$, the results are indeterminate, but it is possible for retarded radiation to be approximately consistent if we only have $p \approx 1$, $f \approx 1$. Specifically, a mixture of the form $(1-\delta)F_{ret} + \delta F_{adv}$ can exist, with

$$\delta \approx \frac{1-p}{1-f} \ll 1 \tag{36}$$

provided $(1-p) \ll (1-f)$. The last condition is not, of course, the same as saying $f \ll p$. For instance if $p=1-10^{-20}$ and $f=1-10^{-10}$, then the above condition is true though not $f \ll p$ in the usual sense. The Einstein-de Sitter model satisfies the condition $p \approx 1$, $f \neq 1$, while the closed Friedmann model satisfies $p \approx 1$, $f \approx 1$, leading to the situation depicted by (36).

## 4  Empirical tests

Is there any way to put to test any of the above considerations? One experiment has been performed, and another announced and later abandoned. Both these tests concern the possible existence of small amounts of advanced radiation, arising from incomplete absorption in the relevant part of the cosmos.

Partridge[30] in 1973 measured the power required to drive a horn antenna which radiated alternately into free space, and into a local ('totally') absorbing cover, which was placed far enough ($20\lambda$) away to eliminate any confusion due to power reflected back. The point of a horn antenna is that it provides a directed beam, which could be pointed towards different parts of the sky. The antenna radiated in the microwave region to ensure that the emergent radiation, when directed towards free space, would not be absorbed by the earth's atmosphere, but could be expected to travel cosmological distances before absorption. Since any change in the power input would be expected to be small, the changes were measured using a phase-sensitive detector to measure the change in the current $\Delta I/I$.

Partridge assumed a relationship of the type

$$P_f = (1 - \delta) P_a, \qquad (37)$$

where $P_f$ denotes the power required while radiating into free space, $P_a$ denotes the power required while radiating into the local absorber, and $\delta$ denotes a small number. Conventionally, what happens to the radiation after it leaves the source should not affect the source in any way, so that we should have $\delta=0$; at any rate we expect $\delta$ to be small, which is the reason for the assumed linear relationship. On the Wheeler-Feynman theory, it is clear from the above discussion that we should have $\delta \geq 0$. Partridge thought this was true of the Hoyle-Narlikar theory as well. The objective of the experiment was to place an upper limit on the value of $\delta$.

For a positive value of $\delta$, as predicted by the Wheeler-Feynman theory, the observed value of $\Delta I/I$ should have been negative for the phase $\phi = 0^\circ$, and positive for the phase $\phi=180^\circ$ of the phase sensitive detector, as compared to the actual observed values $+(14.7\pm7.7)\times10^{-8}$, and $-(10.3\pm7.7)\times10^{-8}$, respectively. I pointed out that[31] the phase sensitive detector is most sensitive for these two phase settings, and both these values correspond to a significant negative value of $\delta$. Such a negative value of $\delta$ is predicted by my theory mentioned above, while the theories of Wheeler-Feynman and Hogarth-Hoyle-Narlikar are already ruled out on grounds of internal and external inconsistency. Partridge actually took a weighted average over various phase settings, to arrive at a mean value of $\delta = (-1.1\pm1.6) \times 10^{-9}$, still negative but no longer significant. Undoubtedly, Partridge's procedure is robust, but this robustness is achieved at the cost of diminished sensitivity of the experiment, since for values of the phase setting other than $0^\circ$ and $180^\circ$ the errors are *systematically* greater.

Partridge concluded that the result was consistent with $\delta=0$, conforming to ordinary expectations, and that either the analysis in terms of the Wheeler-Feynman theory was incorrect or that absorption along the future light cone was better than 1 part in $10^9$.

Pegg subsequently suggested that, in order to obtain a non-null result, it was necessary to use a dynamic absorber, a 'chopper', which would affect past and future absorbers differently. On this basis, an experiment was proposed by Heron and Pegg.[32] The experiment also proposed to exploit the fact that, with a dynamic absorber, the expected change in power input would be time varying. One could use a square wave input, so that any modulation of this, by the 'chopper' could be easily detected. It was estimated that the experiment would, thereby, have an increased sensitivity of 1 part in $10^{12}$, compared to the accuracy of 1 part in $10^9$ in Partridge's experiment.

    The experiment was eventually abandoned due to the absence of a picosecond switch, which now essentially exists.

    Experiments of Partridge's type do not exhaust all empirical possibilities, and we will consider some other possibilities in the next part, including the possibility that a mixture might result in a time which is structured rather than linear.

## 5 Conclusions

The many-body equations of motion, based on retarded interactions, help to resolve the recurrence and reversibility paradoxes of thermodynamics, by destroying the hypotheses underlying those paradoxes. The derivation of the Lorentz-Dirac equation of motion for an electron, and hence the origin of preacceleration, is based on an invalid approximation. The existence of mixed interactions, or a 'tilt in the arrow of time,' is not logically impossible, and may even help to resolve deeper paradoxes. Such a 'tilt' may, in fact, have been detected by Partridge's experiment. Further experiments with increased sensitivity are now possible.

**Appendix: Derivation of the relativistic two-body equations of motion with a tilt in the arrow of time**

We give a summary derivation of the two-body equations of motion. Our aim is to make a clear statement of the physical assumptions used in writing down, say, equation (1) at the beginning of the chapter. It is *not* our intention to simplify the algebra.

The assumptions are the following.
(i) The particles are mathematical points with no size, so that one may speak properly of the world-line of a particle.
(ii) The two charged particles interact electromagnetically. In the retarded case, equation (1), it is assumed that the electromagnetic field (potential) due to a moving charge is that given by the usual (Lienard-Wiechert) retarded potential (equation 2a of Part VA). In the case of a tilt in the arrow of time, one uses mixed potentials (equation 3 of Part VA, with $\alpha$, presumed small, determined from further theoretical and empirical considerations).
(iii) A charged particle moves in accordance with the Heaviside-Lorentz force law. (Equation 4 of Part VA gives the non-relativistic form.)

We partly follow the notation of Synge. Spacetime is taken to be flat, the spacetime coordinates are $x_\mu$ with $x_4 = ict$. This is not the conventional notation, but has the advantage that it avoids the distinction between covariant and contravariant components: all suffixes are written as subscripts. However, we let Greek suffixes range from 1 to 4 and Latin suffixes range from 1 to 3.

The inner product (dot product) of two vectors $V_\mu$, $W_\mu$ is written

$$(VW) = V_\mu W_\mu. \qquad (A.1)$$

For a unit vector, $(VV) = \pm 1$ according as it is space-like or time-like.

We let:

$$\begin{aligned}
L, L' &= \text{world-lines} \\
m, m' &= \text{proper masses} \\
e, e' &= \text{charges (in esu)} \\
\lambda_\mu, \lambda'_\mu &= \text{unit tangents} \\
ds, ds' &= \text{elements of proper time} \\
u_r, u'_r &= \text{velocities} \\
\gamma^{-2} &= 1-u^2/c^2, \quad \gamma'^{-2} = 1-u'^2/c^2 \\
u^2 &= u_r u_r \\
\mu &= \frac{m'}{m}, \quad k = -\frac{ee'}{m'c^2}
\end{aligned} \qquad (A.2)$$

The Lienard-Wiechert retarded potential may be written more elegantly as follows.

$$A_\mu = -\frac{e\,\lambda_\mu}{(\lambda\xi)} \qquad (A.3)$$

This notation, however, suppresses the key fact that the potential $A_\mu$ refers to a point $P'$ on the world line $L'$, while the tangent vector $\lambda_\mu$ refers to a *different* point $P$ on the world-line $L$. The retardation vector $\xi$ is the vector connecting $P$ to $P'$. This fact is made evident in Fig. 5.

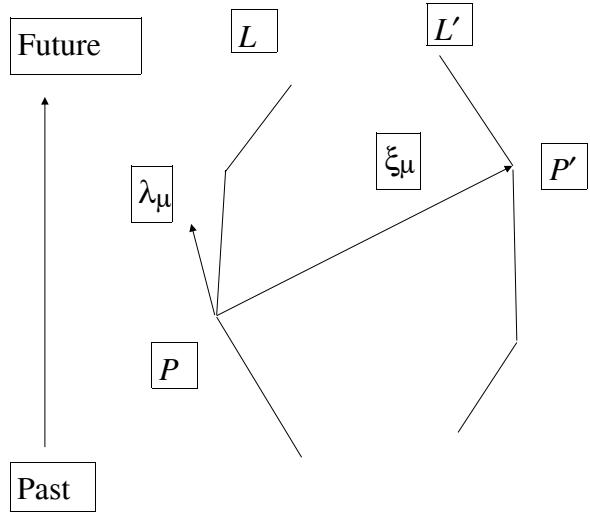

**Fig. 5: The retardation vector**

*Given any point $P'$ on the world line $L'$, to obtain the retarded potential $A_\mu$ at $P'$ due to the world line L, let P be the point at which the backward null cone from $P'$ meets L. Let $\lambda$ be the tangent vector to L at P, and let $\xi$ be the retardation vector joining P to $P'$. Now apply equation (A.3).*

The corresponding electromagnetic field is given by

$$F_{\mu\nu} = \frac{\partial A_\mu}{\partial x'_\nu} - \frac{\partial A_\nu}{\partial x'_\mu} \qquad (A.4)$$

$$= e\,(P_\mu \xi_\nu - P_\nu \xi_\mu), \qquad (A.5)$$

where

$$P_\mu = -\frac{\lambda_\mu}{(\lambda\xi)^3}\left[1 + \left(\frac{d\lambda}{ds}\,\xi\right)\right] + \frac{1}{(\lambda\xi)^2}\frac{d\lambda_\mu}{ds}. \qquad (A.6)$$

The required equations of motion are, therefore, given by

$$m'\frac{d\lambda'_\mu}{ds'} = \frac{e'}{c^2}\,F_{\mu\nu}\,\lambda'_\nu, \qquad (A.7)$$

which reduces to

$$\frac{d\lambda'_\mu}{ds'} = -k\,[P_\mu(\lambda'\xi) - \xi_\mu(\lambda'P)]. \qquad (A.8)$$

Similarly, we obtain the equation of motion for the other particle:

$$\frac{d\lambda_\mu}{ds} = -\mu k \ [P'_\mu(\lambda\xi') - \xi'_\mu(\lambda P')]. \tag{A.9}$$

Equations (A.8) and (A.9) are the generalized forms of equation (1).

The neglect of radiation damping is not in any way essential to the conclusions that follow from equation (1). One can add the radiation damping term to (A.7) to obtain

$$m'\frac{d\lambda'_\mu}{ds'} = \frac{e'}{c^2} F_{\mu\nu} \lambda'_\nu - \frac{2 e'^2}{3 c^2} \left[\frac{d^2\lambda'_\mu}{ds'^2} - \lambda'_\mu \left(\frac{d\lambda'_\nu}{ds'} \frac{d\lambda'_\nu}{ds'}\right)\right] \tag{A.10}$$

It is easy to see that this makes the equation messier: though the highest order derivative now appears in the radiation damping term, this does not change the basic *type* of the equation.

The type of the equation does change if we use advanced or mixed potentials in place of the retarded potential (A.3). Mixed potentials are given by a convex combination of retarded and advanced potentials (equation 3, Part VA):

$$A_\mu = -e \ [\ \alpha \frac{\lambda}{(\lambda\xi)} - (1-\alpha) \frac{\overline{\lambda}}{(\overline{\lambda\xi})} \ ] \tag{A.11}$$

where $Q$ is the point at which the forward null cone from $P'$ intersects the world line $L$, $\overline{\lambda}$ is the tangent vector to $L$ at $Q$, and $\overline{\xi} = \overrightarrow{QP'}$ is the *advance* vector connecting $Q$ to $P'$. Since $\alpha$ is a constant number, the differentiation to obtain the fields may be carried out exactly as in (A.4) and (A.5), and we will not bother to write out explicitly the resulting equations which replace (A.8), (A.9) and (A.10).

We emphasize that a relativistic treatment of this problem is essential. One can certainly conceive the possibility of posing the problem non-relativistically, using the non-relativistic expression for the Heaviside-Lorentz force law (5, Part IIIB). But this would not be a very nice thing to do, because the Heaviside-Lorentz force is velocity-dependent, whereas force is (and ought to be) a Galilean invariant in the framework of Newton's laws. (Hence, the Heaviside-Lorentz force law was a major step towards relativity.) Thus, the key feature of these equations of motion (viz. that they are equations with deviating arguments) is a direct outcome of the postulated finiteness of the speed of interaction.

Secondly, we have seen that the qualitative effects are not confined to relativistic velocities. Consequently, there is a case to use these equations as the basic equations of motion for statistical mechanics.

Finally, we notice that though the field-picture and the particle-picture view the origin of radiation damping differently, this distinction did not enter in any essential way in the equations. This is in line with Poincaré's basic philosophy (expounded in Part IIIB), that the equations are the key, and so long as we do not change the equations it does not matter very much which mental picture we feel comfortable with. The equations, however, do change with the relativistic postulate, and they change further if one postulates a tilt in the arrow of time.

## Notes and References